\documentclass{elsart}
\usepackage{amsmath,amsfonts,amssymb}
\usepackage{bbold,euscript}
\def\ope[#1][#2]{\mathord{#2\over{\ifnum#1=1 {z-w} \else {(z-w)^{#1}}\fi}}}
\def\pair<#1,#2>{\mathop{\left\langle#1\mathbin{,}#2\right\rangle}\nolimits}

\renewcommand{\d}{\partial}

\newcommand{\sG}{\ensuremath{\mathsf G}}
\newcommand{\sT}{\ensuremath{\mathsf T}}
\newcommand{\sJ}{\ensuremath{\mathsf J}}
\newcommand{\sF}{\ensuremath{\mathsf F}}
\newcommand{\sB}{\ensuremath{\mathsf b}}
\newcommand{\sC}{\ensuremath{\mathsf c}}
\newcommand{\SL}{\ensuremath{\mathrm{SL}}}
\newcommand{\1}{\ensuremath{\mathbb 1}}

\newcommand{\R}{\ensuremath{\mathbb R}}
\newcommand{\Z}{\ensuremath{\mathbb Z}}
\newcommand{\II}{\ensuremath{\mathrm {I\!I}}}
\newcommand{\IIA}{\ensuremath{\mathrm {I\!IA}}}
\newcommand{\IIB}{\ensuremath{\mathrm {I\!IB}}}
\newcommand{\eM}{\ensuremath{\EuScript{M}}}
\newcommand{\NPB}[3]{{\sl Nucl. Phys.} {\bf B#1} (#2) #3}
\newcommand{\PRD}[3]{{\sl Phys. Rev.} {\bf D#1} (#2) #3}
\newcommand{\PLB}[3]{{\sl Phys. Lett.} {\bf #1B} (#2) #3}
\newcommand{\MPLA}[3]{{\sl Mod. Phys. Lett.} {\bf A#1} (#2) #3}
\begin{document}
\begin{frontmatter}
\title{F-theory and the universal string theory}
\author[QMW]{Jos\'e~M~Figueroa-O'Farrill
\thanksref{emailjmf}
\thanksref{EPSRC}}
\address[QMW]{Department of Physics, Queen Mary and Westfield College,
Mile End Road, London E1 4NS, UK}
\thanks[emailjmf]{\tt mailto:j.m.figueroa@qmw.ac.uk}
\thanks[EPSRC]{Supported by the EPSRC under contract GR/K57824.}
\begin{abstract}
We apply the techniques of the ``universal string theory'' to the
``manifold'' paradigm for superstring/M-theory and come up with a
candidate manifold: the manifold of F-theory vacua, defined in
conformal field theoretical terms.  It contains the five known
superstring theories as particular vacua; although the natural vacua
are $(10+2)$-dimensional.  As a byproduct, a natural explanation
emerges for the compactness of the extra two coordinates in F-theory.
\end{abstract}
\end{frontmatter}

\section{Introduction}

After the dust of the first superstring revolution settled \cite{GSW}
we were left with five distinct yet perturbatively consistent
superstring theories in ten dimensions.  Some of these theories were
soon shown to be related after compactification; for example, the two
heterotic theories \cite{NSW} and the two type {\II} theories were
shown to be pairwise T-dual \cite{Tduality} after compactifying on a
circle down to nine dimensions.  These duality transformations
preserve the order of the string coupling constant; they can thus be
proven using the techniques of perturbative string theory.  The second
superstring revolution (see \cite{Review} for a progress report) was
ushered in by compelling evidence that some of these theories are
related by duality transformations which invert the string coupling
constant.  For example, type {\IIB} is believed to be self-dual
\cite{SDuality}, whereas the type I theory and the heterotic theory
with group $SO(32)$ are believed to be dual
\cite{WittenStringDynamics}.  Indeed, from these two ``ur-dualities''
one can derive the other known dualities.  One of the biggest
surprises yet that the second superstring revolution had in store is
the evidence of an eleven-dimensional theory (M-theory) whose
compactification down to ten dimensions describe the strongly coupled
type {\IIA} and $E_8\times E_8$ heterotic theories, and whose
effective low-energy physics is described by eleven-dimensional
supergravity \cite{WittenStringDynamics}.  A dense web of dualities
can then be derived by compactification of the string theories or of
M-theory on diverse manifolds thus providing independent checks of the
fundamental assumptions.

The emerging picture is much like that of a manifold for which we are
only given a set of patches of local coordinates and for which we are
told how to change variables on the overlaps.  The analogy is strong:
the perturbative theories are the analogues of local coordinates,
which are only valid in the domain of validity of the perturbation
expansions, and duality transformations are the analogues of the
transition functions at the overlaps.  Yet it is also possible that
this analogy fails at some level: for instance, there is no known
perturbative description for M-theory (indeed there are arguments that
the theory is inherently nonperturbative) and moreover it is not clear
that there is a nonempty overlap between the perturbative
domains---for instance, it is difficult to imagine a region where the
type {\IIB} theory has simultaneous convergent perturbative expansions
in $\lambda_{\IIB}$ and $1/\lambda_{\IIB}$.  Nevertheless I believe we
may learn something by taking the manifold analogy seriously and in
this paper we shall propose a concrete model for this manifold: the
space $\eM$ of F-theory vacua, to be defined below in perturbative
(i.e., CFT) terms.  The manifold $\eM$ has the known superstring
theories as particular vacua, but allows for much more.  In the
process we will give what we believe to be a natural explanation for
the compact nature of the extra two coordinates appearing in F-theory
\cite{FTheory}.  As such this proposal is rather conservative in that
``strings'' still play a privileged role---the manifold is locally
(i.e., perturbatively) described by two-dimensional conformal field
theories.  Notwithstanding the calls for $p$-brane democracy
\cite{pBraneDemocracy} brought about by the key role played by other
extended objects in our understanding of duality, the fact remains
that in the perturbative farm strings are more equal.

Another attractive aspect of the manifold idea is that it allows for
continuous deformations of these theories.  One such deformation could
correspond to changing the expectation value of the ten-dimensional
dilaton.  In the case of say the type {\IIB} theory there are
expectation values of the dilaton (near zero) where type {\IIB}
perturbation theory is invalid.  Duality would tell us that in this
regime the dual theory should have a valid perturbative description,
but the dual theory is again type \IIB.  Thus there seem to be gaps
which the present proposal might fill, since it allows in principle
for the existence of other perturbative ``superstring'' theories.  In
the case of the type {\IIA} or the $E_8\times E_8$ heterotic strings,
increasing the vacuum expectation value of the dilaton should give us
an eleven-dimensional vacuum (one of whose dimensions is compact of
size proportional to the logarithm of the vacuum expectation value of
the dilaton).  We have so far been unable to find such vacua in this
manifold; but its existence is not ruled out: only that its explicit
description has so far eluded us.  Finding such a vacuum seems to be
the litmus test of the present proposal or, indeed, of any other such
proposal.

The plan of this paper is as follows.  In the next section we use the
ideas of ``universal string theory'' in order to bring the five
ten-dimensional superstring theories together into the same manifold
of vacua.  That this is possible is implicit in some of the earlier
work in this topic, although it never really appeared in print.  The
new ingredient is the self-duality of the type {\IIB} superstring,
which will make us modify the manifold of vacua to take into account
the new enlarged gauge symmetry present in the type {\IIB} D-string.
This is described in Section 3, where we will also discover a natural
explanation for the compactness of the extra two dimensions of
F-theory.  In Section 4 we briefly compare this proposal with recent
work along similar lines and conclude with some speculative comments
about the role of conformal field theory in nonperturbative string
theory.

\section{Universal string theory}

We start by reviewing some of the philosophy behind the ``universal
string theory'' and how this helps to put the five ten-dimensional
superstring theories under the same footing; that is, as points of the
same manifold of vacua.  Already the two type {\II} superstrings and
(after a quotient which identifies left- and right-movers) the type I
superstring can be understood as different points in the same manifold
of vacua: namely, the manifold of $N{=}(1,1)$ superconformal field
theories with central charges $(c_L,c_R) {=} (15,15)$.  Let us call
this manifold $\eM_0$.  On the other hand, the two heterotic string
theories are points in the manifold of $N{=}(1,0)$ superconformal
field theories with central charges $(15,26)$.  The original
observation of ``universal string theory''
\cite{BerkovitsVafa,FUST,IshikawaKato} is that any bosonic string
theory can be understood as an $N{=}1$ superstring theory.  (Other
embeddings are possible and we shall comment on some of them in due
course.)  In fact, this works purely at the level of meromorphic
conformal field theory and hence separately on the left and on the
right.  This means that we can also think of the heterotic strings as
particular vacua in $\eM_0$.

Let us briefly go through the details.  We use the standard
conventions of operator product expansions and normal-ordered
products.  Let $T(z)$ be the holomorphic component of an energy
momentum tensor with central charge $26$.  Introduce a fermionic BC
system $(\sB,\sC)$ of conformal weights $(\tfrac{3}{2},-\half)$ and
define
\begin{align}
\sG &= \sB + \sC T + \sC\d\sC\sB + \tfrac{5}{2} \d^2\sC\notag\\
\sT &= T - \tfrac{3}{2} \sB\d\sC - \half \d\sB\sC + \half
\d^2(\sC\d\sC)~.\label{eq:embedding}
\end{align}
Then $\sG$ and $\sT$ obey an $N{=}1$ superconformal algebra with
central charge $15$.  This means that we can compute both the BRST
cohomology of the bosonic string defined by $T$ and that of the
superstring defined by $(\sG,\sT)$.  As is well-known they are
isomorphic; more precisely, the BRST cohomology of the superstring
$(\sG,\sT)$ at a {\em fixed\/} picture is isomorphic with the BRST
cohomology of the bosonic string $T$.  Moreover the GSO projector is
the identity on the cohomology, so the physical spectra agree.
As an example, we can take $T$ to be the following:
\begin{equation}
T = -\half \eta_{\mu\nu} \d X^\mu \d X^\nu - \half \delta_{ij}\d Y^i
\d Y^j~,\label{eq:heterotic}
\end{equation}
where $X^\mu$, $\mu=0,\ldots,9$ coordinatise the embedding of the
worldsheet in ten-dimensional Minkowski space, and $Y^i$,
$i=1,\ldots,16$ coordinatise a sixteen-dimensional torus
$\R^{16}/\Lambda$ with $\Lambda$ one of the two even self-dual
euclidean lattices of rank sixteen.  This corresponds to the
left-moving, say, energy-momentum tensor of the heterotic string
corresponding to $\Lambda$.  Therefore the (1,1) superconformal field
theory obtained by gluing the right-moving sector of the
ten-dimensional superstring with the above left-moving superconformal
field theory with generators $(\sT,\sG)$ given by \eqref{eq:embedding}
and \eqref{eq:heterotic} has the same physics as the corresponding
heterotic string.

We therefore have all superstring theories as different points in the
same manifold; however we will now argue that this cannot be the end
of the story.  The missing ingredient is provided by the self-duality
of the type {\IIB} string, which will force us to rethink the gauge
principle on the worldsheet.  This is the same argument used by Vafa
to introduce F-theory \cite{FTheory}.

\section{Incorporating the self-duality of the type {\IIB} superstring}

The massless spectrum of the type {\IIB} string contains a two-form in
the R-R sector, which couples to a string---the D-string.  Under
S-duality \cite{SDuality}, this two-form gets interchanged with the
two-form in the NS-NS sector, and hence the fundamental string (the
F-string) and the D-string are switched.  This implies that the
strongly coupled physics of the type {\IIB} string possesses a
perturbative description where the D-string is now the fundamental
object.  All evidence points to the fact that this string is again of
type \IIB.  However, {\it a prima facie}, the F-string and the
D-string look different.  From its interpretation as a D-brane, open
strings can end on the worldsheet of the D-string and the massless
vector in the open string spectrum induce a $U(1)$ gauge field there.
Hence the gauge principles of the F- and D-strings differ.  Of course,
from what we have learned about the ``universal string theory'' this
is no longer a cause for concern.  Indeed it is possible to extend the
gauge principle of the type {\IIB} string to include a gauged $U(1)$
current without changing the spectrum.  This was done in
\cite{FTheory} as the main motivation behind F-theory.  Here we will
review it in the light of the ``universal string theory,'' and as a
corollary it will follow that the extra two coordinates in F-theory
are compact -- an otherwise {\em ad hoc\/} assumption.

\subsection{Enlarging the gauge principle}

Let us first consider a bosonic string background.  This is a theory
of two-dimensional gravity coupled to some matter.  The graviton
couples to the energy-momentum tensor $\sT$.  If we now add a $U(1)$
gauge field, it will couple to a vector $\sJ$.  We therefore would
like to investigate under which conditions the algebra generated by
$T$ and $\sJ$ can be used consistently to define a (generalised)
string theory.  We consider one chiral sector.  The operator product
algebra satisfied by $\sT$ and $\sJ$ has two independent parameters:
the central charge and the constant $a$ in the double pole
$[\sJ,\sJ]_2$.  The algebra is given by
\begin{xalignat}{3}
[\sT,\sT]_4 &= \half c\1 & \quad [\sT,\sT]_2 &= 2 \sT & \quad
[\sT,\sT]_1 &= \d\sT\notag\\
[\sT,\sJ]_2 &= \sJ & \quad [\sT,\sJ]_1 &= \d\sJ&& \notag\\
[\sJ,\sJ]_2 &= a\1 &&&&\label{eq:AtiyahAlgebra}
\end{xalignat}
Of course, there are only two discrete choices for the parameter $a$:
it is either zero or it can be rescaled to any convenient nonzero
value.  To define the BRST current we introduce fermionic ghost pairs
$(b,c)$ and $(\sB,\sC)$ of weights $(2,-1)$ and $(1,0)$ respectively.
The BRST current is then given by
\begin{equation}
j = c\sT + \sC\sJ + bc\d c - c\sB\d\sC~,\label{eq:ABRST}
\end{equation}
and it is routine to verify that the BRST operator $d = [j,-]_1$
squares to zero provided that $a{=}0$ and $c{=}28$.

A particular realisation of this algebra is provided by a bosonic
string propagating in a 28-dimensional pseudo-euclidean space with
signature $(26,2)$.  The signature can be understood from unitarity
or, alternatively, in the heterotic context at least, from modular
invariance.  Choose a null vector $v$ and define
\begin{equation}
\sT = -\half \d X \cdot \d X\quad\text{and}\quad \sJ = v
\cdot \d X~.
\end{equation}
Then $\sT$ and $\sJ$ obey \eqref{eq:AtiyahAlgebra}.  It was argued in
\cite{OoguriVafa} that the BRST cohomology of this system agrees with
that of an underlying bosonic string propagating in a 26-dimensional
Minkowski subspace perpendicular to $v$ and not containing $v$,
provided that we identify states whose momenta differ by a multiple of
$v$.  More generally, let $T$, with central charge 26, describe the
background of a bosonic string theory.  Add two extra free bosons $\d
X^\pm$ with operator product expansion $[\d X^+,\d X^-]_2 = \1$.  Then
define
\begin{equation}
\sT = T + \d X^+ \d X^-\quad\text{and}\quad \sJ = \d
X^-~.\label{eq:ovaa}
\end{equation}
This forms a realisation of the algebra \eqref{eq:AtiyahAlgebra} whose
BRST cohomology can be shown to be isomorphic to that of the
bosonic string with background $T$ provided one identifies all states
regardless of their momentum in the $X^-$ direction.

This identification of states is reminiscent of the notion of
``pictures'' familiar from the BRST quantisation of the NSR string.
There, the bosonic ghosts corresponding to the supercurrent have a
(countably) infinite number of inequivalent vacua which can be
understood as the momenta in one of two auxiliary compactified
dimensions introduced by the bosonisation procedure.  The picture
changing operator interpolates between these different vacua,
commuting with the BRST operator and thus introducing an infinite
degeneracy in the cohomology.  In fact, as we now explain, these two
phenomena are not unrelated.

In the formalism of the ``universal string theory,'' the way one goes
about enlarging the gauge symmetry of a bosonic string theory with
background $T$ is to add a BC system of equal conformal properties but
opposite statistics to the ghosts for the new generators we hope to
introduce.  This guarantees that, taking ghosts into account, the
total central charge remains zero.  In the case of interest, this
means introducing a bosonic BC system $(\Hat\beta,\Hat\gamma)$ with
weights $(1,0)$.  Then in the chiral algebra generated by $T$,
$\Hat\beta$ and $\Hat\gamma$ we should find a realisation of the
algebra \eqref{eq:AtiyahAlgebra}.  This is most easily achieved by
bosonising $(\Hat\beta,\Hat\gamma)$ \cite{FMS}.  The conformal field
theory of a bosonic BC system can be embedded in that of two free
bosons compactified on a torus with lorentzian signature $(1,1)$.
Call these bosons $\d X^\pm$.  Then a realisation of the algebra
\eqref{eq:AtiyahAlgebra} is given by a deformation of \eqref{eq:ovaa}:
\begin{equation}
\sT = T + \d X^+ \d X^- + \half \d^2X^-\quad\text{and}\quad
\sJ = \d X^-~.\label{eq:ustaa}
\end{equation}
The choice of deformation for $\sT$ is such that $\sT = T -
\Hat\beta\d\Hat\gamma$, where the second term is the energy-momentum
tensor for the bosonic BC system.  In terms of $X^\pm$, the bosonic BC
fields are given by
\begin{equation}
\Hat\beta = -\half e^{-X^+} (\d X^+ + \d X^-) \quad\text{and}\quad
\Hat\gamma = e^{X^+}~.\label{eq:bosonisation}
\end{equation}
Using the same techniques as in \cite{OoguriVafa} one can show that
the BRST cohomology of this generalised string theory at a fixed value
of $p^+$ agrees with the cohomology of the underlying bosonic string.
The operator which shifts $p^+$ is the ``spectral flow'' of the null
current $\exp (q X^-)$; but this operator also changes the pictures of
the BC system.

Let us close this discussion by remarking that picture-changing has a
spacetime interpretation in this context.  Suppose that $T$ is the
energy-momentum tensor of a critical bosonic string propagating in
26-dimensional Minkowski spacetime.  Then $\sT$ corresponds to the
string propagating on a $(26+2)$-dimensional pseudo-euclidean space.
The BRST operator is invariant under the subgroup of the $(26+2)$
pseudo-euclidean group of motions which preserves the null vector $v$.
This is nothing but the $(25+1)$ conformal group, which does not act
linearly in Minkowski spacetime but does on the larger space.
Symmetries of the BRST operator induce symmetries in the cohomology,
hence we would expect that the spectrum should assemble itself into
representations on the conformal group.  We know that the physical
spectrum of the bosonic string only possesses $(25+1)$ Poincar\'e
covariance, so what happens to the special conformal transformations?
They are such that they necessarily shift $p^+$, whence they change
the picture.  By definition a picture-changing operator is a BRST
invariant operator which changes the picture, whence the special
conformal transformations are picture-changing operators.

A remarkable fact of this treatment is that the appearance of the
lorentzian torus is very natural.  In other words, by enhancing the
gauge principle on the worldsheet to incorporate the extra $U(1)$
gauge invariance we are forced to reinterpret bosonic string vacua
corresponding to propagation on a given manifold $M$, as propagation
in a manifold which at least locally is of the form $M\times T^2$
where $T^2$ is the lorentzian torus corresponding to the bosons
$X^\pm$.  This theory is precisely the F-theory introduced in
\cite{FTheory}, except that there the compactness of the extra two
coordinates was an {\em ad hoc\/} assumption.

\subsection{The manifold of F-theory vacua}

The considerations in the previous subsection can be
supersymmetrised.  In this way we will be able to think of all
superstring theories (and indeed any point in the manifold $\eM_0$ of
vacua described above) as belonging to a larger space $\eM$: the
manifold of F-theory vacua to be defined presently.

We defined $\eM_0$ to be the space of $N=(1,1)$ superconformal field
theories with central charges $(15,15)$.  We will now apply the
methods of ``universal string theory'' to once again enlarge the gauge
symmetry.  The symmetry algebra we are after is the
supersymmetrisation of \eqref{eq:AtiyahAlgebra}, with generators
$\sT$, $\sG$, $\sJ$ and $\sF$, of weights $2$, $\tfrac{3}{2}$, $1$ and
$\half$ satisfying the following operator products in addition
to those in \eqref{eq:AtiyahAlgebra} with $c{=}18$ and
$a{=}0$:
\begin{xalignat}{2}
[\sG,\sG]_3 &= 12 \1 & \quad [\sG,\sG]_1 &= 2 \sT \notag\\
[\sT,\sG]_2 &= \tfrac{3}{2} \sG & \quad [\sT,\sG]_1 &= 2 \d\sG\notag\\
[\sG,\sJ]_2 &= \sF & \quad [\sG,\sJ]_1 &= \d\sF\notag\\
[\sG,\sF]_1 &= \sJ &&\notag\\
[\sT,\sF]_2 &= \half\sF & \quad [\sT,\sF]_1 &=
\d\sF~.\label{eq:SuperAtiyahAlgebra}
\end{xalignat}
We have chosen the above values of $c$ and $a$ in order to be able to
gauge this algebra; that is, for it to admit a BRST operator.  We can
interpret such a conformal field theory as a superstring propagating
on a ($10{+}2$)-dimensional spacetime.  Indeed a realisation of this
algebra is given by:
\begin{xalignat*}{3}
\sT &= -\half \d X \cdot \d X - \half \psi \cdot \d\psi & \quad \sJ
&= v \cdot \d X\\
\sG &= \psi \cdot \d X & \quad \sF &= v \cdot \psi~,
\end{xalignat*}
where $v$ is a null vector in a pseudo-euclidean space of signature
$(10,2)$.  We can construct a larger class of realisations in the
following way.  Let $T$ and $G$ obey the superconformal algebra with
central charge $c{=}15$.  We introduce now two free bosons $\d X^\pm$
and also two free fermions $\psi^\pm$ with operator product expansion
$[\d X^+,\d X^-]_2=\1$ and $[\psi^+,\psi^-]_1=\1$.  The fields
\begin{xalignat}{3}
\sT &= T + \d X^+ \d X^- -\half \psi^+ \d\psi^- - \half\psi^-\d\psi^+
& \quad \sJ &= \d X^-\notag\\
\sG &= G + \psi^+ \d X^- + \psi^-\d X^+ & \quad \sF &= \psi^-
\label{eq:10+2}
\end{xalignat}
obey the operator product expansions \eqref{eq:SuperAtiyahAlgebra}.

Let us define $\eM$ to be the space of F-theory vacua, understood as
conformal field theories admitting a realisation of the algebra
\eqref{eq:SuperAtiyahAlgebra}.  We will now show that it contains
$\eM_0$ as a submanifold, in the sense that every superstring vacuum
can be understood as an F-theory vacuum.  To this effect let $T$ and
$G$ define a point in $\eM_0$; that is, a superstring background.  We
introduce BC systems to cancel the ghosts for the new generators of
the enhanced symmetry.  In this case we will have new generators $\sJ$
and $\sF$, whence we will have to introduce a bosonic BC system
$(\Hat\beta,\Hat\gamma)$ of weights $(1,0)$ and a fermionic BC system
$(\psi^+,\psi^-)$ of weights $(\half,\half)$.  Of course, this latter
system is nothing but a pair of free fermions.  We now bosonise the
former BC system in terms of $\d X^\pm$ as in \eqref{eq:bosonisation}.
Let us define:
\begin{xalignat*}{2}
\sT &= T + \d X^+ \d X^- + \half \d^2X^- - \half \psi^+\d\psi^- -
\half \psi^-\d\psi^+ & \quad \sJ &= \d X^-\\
\sG &= G + \psi^+\d X^- + \psi^- \d X^+ + \d\psi^- & \quad \sF &=
\psi^-~,
\end{xalignat*}
which is a deformation of \eqref{eq:10+2} chosen so that $\sT$ and
$\sG$ are the total energy momentum tensor and total supercurrent,
including that of the two BC systems.  Analogous techniques to those
in \cite{OoguriVafa} allows us once more to show that the BRST
cohomology of this generalised superstring theory at a fixed value of
$p^+$ agrees with the cohomology of the underlying superstring.  The
comments made above on the picture-changing again apply.

In summary, we have shown that there is a space $\eM$ of certain
superconformal field theories, which contains all known superstring
theories and which incorporates the self-duality of the type {\IIB}
string in the spirit of F-theory.  In fact, $\eM$ can be understood as
the space of F-theory vacua, which does not just contain the type
{\IIB} superstring vacua, but has indeed been seen to contain the vacua
of all the superstring theories, including the heterotic ones.  The
physical spectrum of a theory corresponding to a point in $\eM$ is
given by the cohomology of the BRST operator for the algebra
\eqref{eq:SuperAtiyahAlgebra}.  It is an open problem to find a point
in $\eM$ whose spectrum exhibits eleven-dimensional Poincar\'
covariance; however we expect that such points exist since the type
{\IIA} and heterotic $E_8\times E_8$ theories correspond to points in
$\eM$ and their strong coupling limit is conjectured to correspond
to compactifications of M-theory in the decompactification limit.

\section{Conclusions and outlook}

We have taken seriously the ``manifold'' paradigm for
superstring/M-theory and proposed as a candidate space, the manifold
$\eM$ of F-theory vacua.  We have done so by the systematic use of the
``universal string theory'' philosophy (see \cite{FMech} for some
general statements).  As a byproduct, an explanation emerges for the
compactness of the extra two dimensions in F-theory, a basic
assumption of \cite{FTheory}.

Two other proposals for $\eM$ have appeared in the recent literature.
In the work of \cite{BGR} ``universal string theory'' ideas are used
to view superstrings as strings with higher world-sheet supersymmetry
and then move about the space of such theories via marginal
deformations.  In particular $\eM$ appears to be the space of
$N{=}(4,4)$ strings whose natural target is zero-dimensional.  In the
work of \cite{KutasovMartinec}, $\eM$ appears as the space of
heterotic $N=(2,1)$ superstrings.  These string theories have
backgrounds whose target spaces correspond to the worldsheet or
worldvolume of strings and membranes.  Applying the machinery of
``universal string theory'' one can understand these theories as
$N{=}(4,4)$ theories and make contact with the work of \cite{BGR}.
Neither of these proposals has so-far exhibit a point with an
effective eleven-dimensional spacetime theory.

Finally, we comment on the role of conformal field theory in
nonperturbative string theory.  With the advent of the second
superstring revolution, the focus has moved away from the worldsheet
and towards the spacetime approach to string theory.  In particular
taking into account the pivotal role played by other extended objects
in our understanding of duality, one may wonder whether conformal
field theory can teach us anything about nonperturbative string
theory.  The present paper and some recent work
\cite{Dolan,BGR,KutasovMartinec,Polyakov} suggest that it can.  In
particular, the intriguing paper of D~Polyakov \cite{Polyakov} points
out a curious connection between picture-changing and S-duality which
resonates strongly with the results in this paper.  Polyakov observes
that the spacetime supersymmetry algebra (in type {\IIB}) depends on
the picture.  Roughly speaking, picture-changing maps a configuration
with no 5-form charge to one with it.  According to the self-duality
hypothesis for the type {\IIB} string, these configurations are
supposed to be $\SL(2,\Z)$-related.  It is therefore tempting to
speculate that this information is somehow encoded in the conformal
field theory BRST formalism of the superstring and in particular in
the picture-changing phenomenon.  This speculation gathers strength
when we notice that in the F-theory context, the $\SL(2,\Z)$ duality
group is geometrised as the modular group acting on the extra torus
via large diffeomorphisms; whereas in the present context, the
picture-changing phenomenon (of the fields which bosonise into the
coordinates of the torus) also has a spacetime interpretation.  I hope
to turn to a more systematic analysis of this problem in a future
publication.

\begin{ack}
It is a pleasure to thank Bobby Acharya, Jerome Gauntlett, Chris Hull
and especially Bill Spence for discussions and suggestions.  I am also
grateful to Warren Siegel for an illuminating e-conversation.
\end{ack}

\end{document}